\documentclass[12pt]{article} 
\usepackage[top=2cm,bottom=2cm,left=3cm,right=3cm]{geometry}
\usepackage{amsfonts}
\usepackage{mathptmx}
\usepackage{amsmath}
\usepackage{mathrsfs}
\usepackage{amsmath}
\usepackage{bm}
\usepackage{fancyhdr}
\usepackage[colorlinks,
citecolor=blue,linkcolor=red]{hyperref}
\usepackage{amsfonts}
\usepackage{natbib}
\usepackage{amsthm}
\usepackage{amssymb}
\usepackage{color}
\usepackage{lscape}
\usepackage{rotating}
\usepackage{subfigure}
\usepackage{graphicx}
\usepackage{placeins}
\usepackage{multirow}
\usepackage{tabularx,booktabs,caption}
\newcolumntype{C}[1]{>{\Centering}m{#1}}

\makeatletter
\AtBeginDocument{%
  \expandafter\renewcommand\expandafter\subsection\expandafter
    {\expandafter\@fb@secFB\subsection}%
  \newcommand\@fb@secFB{\FloatBarrier
    \gdef\@fb@afterHHook{\@fb@topbarrier \gdef\@fb@afterHHook{}}}%
  \g@addto@macro\@afterheading{\@fb@afterHHook}%
  \gdef\@fb@afterHHook{}%
}
\makeatother

\renewcommand{\hat}{\widehat}

\usepackage{enumerate}
\usepackage{float}
\usepackage{amssymb}

\newcommand{\E}{\mathbb{E}}
\renewcommand{\P}{\mathbb{P}}
\renewcommand{\hat}{\widehat}

\newcommand{\R}{\mathbb{R}}

\renewcommand\arraystretch{1.1}
\renewcommand{\theequation}{\thesection.\arabic{equation}}

\usepackage{epstopdf}
\usepackage{enumitem}
\usepackage{multirow}
\usepackage{booktabs}

\newtheorem{theorem}{Theorem}

\newtheorem{lemma}{Lemma}

\theoremstyle{definition}

\newtheorem{example}{Example}
\newtheorem{remark}{Remark}
\newtheorem{assumption}{Assumption}

\begin{document}

\title{\vspace{-4cm} \bf Statistical inference in  massive datasets by  empirical likelihood  }

\author{
~~ Xuejun MA \thanks{School of Mathematical Sciences, Soochow University, 215006, Suzhou, China,  stamax360@outlook.com}
~~ Shaochen WANG \thanks{School of Mathematics,
 South China University of Technology, Guangzhou, 510640, P.R. China. mascwang@scut.edu.cn}
~~Wang ZHOU \thanks{Department of Statistics and Applied Probability, National University of Singapore,  6 Science Drive 2, 117546, Singapore. stazw@nus.edu.sg}
}

\date{}
\maketitle

\begin{abstract}
In this paper, we propose a new statistical inference method for massive data sets, which is very simple and efficient by combining  divide-and-conquer method and empirical likelihood. Compared with two popular methods (the bag of little bootstrap and the subsampled double bootstrap), we make full use of data sets, and reduce the computation burden. Extensive numerical studies and real data analysis demonstrate the effectiveness and flexibility of our proposed method. Furthermore, the asymptotic property of our method is derived.
\end{abstract}

\begin{quote}
\noindent
{\sl Keywords}: Bootstrap; divide-and-conquer; hypothesis test; empirical likelihood.
\end{quote}

\begin{quote}
\noindent
{\sl MSC2010 subject classifications}: Primary 62G10; secondary 62G05.
\end{quote}

\thispagestyle{empty}
\pagenumbering{gobble}

\newpage
\pagenumbering{arabic}

\setcounter{page}{1}

\section{Introduction}

With the rapid development of science and technologies, massive data can be collected at a large speed, especially in internet and financial fields. It is generally recognized that two major challenges in large-scale learning are estimation and inference due to large amount of computation.

For statistical inference on massive data sets, \cite{Kleiner2014} proposed the bag of little bootstrap (BLB) to assess the quality of estimators. However, they used only a small number of random subsets, and partial observations from each subset. This implies less efficiency in application. So, \cite{Sengupta2016} developed the subsampled double bootstrap (SDB) method which not noly saves cost computation, but also takes more information of full data than BLB. Compared with the traditional bootstrap (TB), BLB and SDB save the computation cost. However, BLB and SDB have some disadvantages. Similar to  traditional bootstrap, they still sample from full dataset, and repeat the whole process many times.  The computational cost is still expensive. On the other hand, they do not use the full data since about 63\% of data points are contained in each resample.

In addition, \cite{Wang2018} proposed  subsampling method to make inference for Logistic regression.  Subsampling method was first proposed by \cite{Ma2015} for linear regression. Generally speaking, it is a two-step subsampling algorithm.  The first step is to get the weight of each data point. In the second step, the weighted estimator is obtained by combining  resample subset with  subsampling weights. In order to get the optimal subsampling strategy, \cite{Wang2018} suggested two methods, minimum mean squared error (mMSE) and  minimum variance-covariance (mVC).  These methods make use of partial data, and rely on the weighted subsampling estimation. Although their efficiency of estimation is high, but their inference does not works well since the subsampling method aims at estimator in nature. Furthermore, one has to estimate the variance-covariance matrix.

In this paper, we propose combining divide-and-conquer (DAC) and empirical likelihood (EL). As we know, DAC is a very effective estimation method for massive data. Firstly, it split entire datasets into $K$ subsets, and each subset is analyzed separately. Secondly, we combine all subset results via average.  \cite{Chen2014} called it ``split-and-conquer", and applied it to the generalized linear model with sparse structure. \cite{Shi2018}) studied the M-estimators with cubic rate of convergence by DAC, and proved that its  convergence rate  is faster than the original M-estimator. We also refer to \cite{Zhang2013}. On the other hand, EL (\cite{Owen1988,Owen1990,Owen2001}) is a powerful nonparametric method to make inference on parameters of population without assuming the form of the underlying distribution, such as mean, quantiles and regression parameters. We will take advantage of DAC and EL. Compared with BLB and SDB, we not only take full data information, but also save the cost computation.  Our method is very simple and efficient. It has two steps. In the first step, we split the sample into random subsets and the estimate of each subset is obtained. In the second step, the estimates are regarded as one sample from a population so that one can apply EL to this simplified sample.

The rest of this article is organized as follows. In Section \ref{sec2}, we explain our method in details, and establish its theoretical property. In Section \ref{sec3},  we assess  the finite sample performance of proposed method via Monte Carlo simulations. A  real data set is analyzed in  Section \ref{sec4}. 
All technical proofs of main results are postponed to Appendix.

\section{Methodology}\label{sec2}
Let $\mathcal{X}_{n}=\{X_{1}, \dots, X_{n}\}$ be a sample consisting of independent and identically distributed observations form some unknown $q$ dimensional distribution $F$.  The parameter of interest is $\theta=\theta(F)\in \R^{p}$. Its estimator is $\hat{\theta}_{n}=\hat{\theta}(\mathcal{X}_{n})$, which could be maximum likelihood estimator, M-estimator, sample correlation coefficient, U-statistics and many others. In this paper, we mainly focus on the inference of $\theta$. Here is our method.

 We first divide the full data set  into $K$ blocks randomly, say $\mathcal{X}_{1n_{1}},\dots, \mathcal{X}_{Kn_{K}}$, and then compute $\{ \hat{\theta}_{1n_{1}}=\hat{\theta}(\mathcal{X}_{1n_{1}}), \dots  \hat{\theta}_{Kn_{K}}=\hat{\theta}(\mathcal{X}_{Kn_{K}})\}$.  For simplicity, we assume $n_{j}=m$ for all $1\leq j\leq K$. The DAC estimator is defined by
 $$
 \widetilde{\theta}_{n}=\frac{1}{K}\sum_{j=1}^{K} \hat{\theta}_{jm}.
 $$

 Now, we discuss the asymptotic properties of  $\widetilde{\theta}_n$. We assume that $p$ and $q$ are fixed and $K, m \to \infty$. Besides, we need the following assumptions.

\begin{assumption}\label{assumption1}
$$
\sqrt{m}(\hat{\theta}_{km}- \theta) = \frac{1}{\sqrt{m}}\sum_{i=1}^{m}\eta_{ki}+R_{k m},\quad k=1, \dots, K,
$$
 where $\eta_{ki}=(\eta_{ki1},\cdots,\eta_{kip})^\top$ and $R_{km}=(R_{km1},\cdots,R_{kmp})^\top$. Here $\eta_{k1}, \dots, \eta_{km}$  are independent and identically distributed vectors with zero mean,  non-singular covariance matrix $\Sigma$ and $\E\|\eta_{k1}\|^4<\infty$. $R_{km}$ are the remainder terms, which satisfy $R_{km}=o_{p}(1)$.
\end{assumption}

\begin{assumption}\label{assumption2}
\begin{enumerate}
  \item[A2.1] $R_n:=\frac{1}{\sqrt{K}}\sum_{k=1}^{K} R_{km} =o_{p}(1)$.
  \item[A2.2] $\max_{1\leq k\leq K} \|  R_{km}\|=o_{p}(m^{-\alpha})$ for some $\alpha>0$.
  \item[A2.3] $K=O(m^{4\alpha})$.
\end{enumerate}
\end{assumption}
Assumption \ref{assumption1} is a commonly used condition.
This is the Bahadur representation of $\hat{\theta}_{n}$, which has very rich literatures. For example,
 \cite{He1996} studied the  Bahadur representations for a general class of M-estimators. \cite{Arcones1996} explored the  Bahadur representation of $L_{p}$ regression estimators. Assumption \ref{assumption2} is about the rate convergence of the remainder term in the Bahadur representation, i.e.,
  It implies  that
$$
\sqrt{n}(\widetilde{\theta}_{n}- \theta) = \frac{1}{\sqrt{n}}\sum_{k=1}^K\sum_{i=1}^{n}\eta_{ki}+R_{n}.
$$
This is a very mild condition.

\begin{theorem}\label{theorem1} Under Assumptions \ref{assumption1}--\ref{assumption2}, we have
  \begin{equation}
    \sqrt{n}\Big( \widetilde{\theta}_n -\theta \Big)\stackrel{d}{\longrightarrow}N(0,\Sigma),
  \end{equation}
as  $m, K\to \infty$,  where $\stackrel{d}{\longrightarrow}$ denotes  convergence in distribution.
\end{theorem}

Theorem \ref{theorem1} implies that if the usual estimator based on the whole sample has the asymptotic normal distribution, the DAC estimator $ \widetilde{\theta}_n$ has the same asymptotic distribution. However, the covariance matrix $\Sigma$ is usually unknown. One has to estimate it first when applying Theorem \ref{theorem1} to make further statistical inference. Sometimes its estimator is hardly obtained. So we propose to use EL as follows.

Since the blocks are disjoint,  $\hat{\theta}_{1m}, \dots, \hat{\theta}_{Km}$ are independent. We can  regard them  as one sample and apply EL to make inference on $\theta$. For notational convenience, let $Y_{km}={\sqrt{m}}\hat{\theta}_{k m}$ and $\mu=\sqrt{m}\theta$. Hence, the empirical likelihood ratio for $\mu$ is given by
\begin{equation}\label{eq20}
 \mathcal{R}(\mu)=\max\left\{   \prod_{k=1}^{K}K\omega_{k} ~\Big|~~\sum_{k=1}^{K}\omega_{k}Y_{k m}= \mu, \omega_{k}\geq 0,\quad \sum_{k=1}^{K}\omega_{k}=1          \right\}.
\end{equation}
By the Lagrange multipliers method, we can find the maximum point
\begin{equation*}
 \omega_{k}= \frac{1}{K} \frac{1}{1 + \lambda^\top(Y_{km}- \mu)},
\end{equation*}
where $\lambda=\lambda(\mu)$ satisfies the equation given by
\begin{equation}\label{eq24}
  0=\frac{1}{K}\sum_{k=1}^{K} \frac{Y_{km}- \mu}{1 + \lambda^\top(Y_{km}- \mu)}.
\end{equation}
As in \cite{Owen1990}, we can get the follow Wilks' theorem.

\begin{theorem}\label{theorem2}
 Under Assumptions \ref{assumption1}--\ref{assumption2}, we have
$$
-2 \log \mathcal{R}(\mu) \stackrel{d}{\longrightarrow} \chi^{2}_{p}
$$
as $K, m\to \infty$.
\end{theorem}

\begin{remark}
 The accuracy of each block estimator increases as $m$ increases. The power of EL increases as $K$  becomes greater. So there is a trade-off between $K$ and $m$. But we are studying massive data, $K$ and $m$ are large enough to guarantee the accuracy of each step's inference. In simulations, we set $n=10^5$, $K=\{50, 100, 150\}$. The numerical results show that our proposed method is not sensitive to $K$.
\end{remark}

 Compared with the BLB and SDB, our method provides a specific asymptotic distribution to make inference on $\theta$. It is unnecessary to apply bootstrap to specify critical values. This reduces the computation burden a lot.

Now, we discuss the computational times of our proposed method, BLB and SDB. Let $t(m)$ be the computational time to estimate $\hat{\theta}_{m}$ based on a sample of size $m$. $c(K)$ denotes the cost time of EL based on $K$ blocks. Table \ref{table1} presents the comparison. In Table \ref{table1}, the column ``Estimation time" means the corresponding time measured in second when one runs Case 1 of Example \ref{example2} in Section \ref{sec3}. As for the other notation, $b$ is the subset size, $S$ is the number of subsets, $R$ is the number of sampled subsets. The detailed setting is shown in Section \ref{sec3}. We run R language with version 3.5.2 in the desktop computer with Intel(R) Core(TM)CPU  i7-4770\@ 3.40GHz processor and 16.0GB RAM. Here we select $b$ of BLB and SDB  to be a litle big so that most information of data can be used. From Table \ref{table1}, one can see that our method reduces the computation burden a lot.

\begin{table}[htbp!]
\renewcommand\arraystretch{1.2}
\centering
\caption{The computational time for different methods. }{%
\begin{tabular}{cccc}
 \\\toprule
Method	&	Cost time &  & Estimation time (seconds) 	\\\midrule
BLB     &  $R\times S\times t(b)$ & $b=n^{0.6}$ &  26.528  \\
         &   & $b=n^{0.8}$ &   209.810 \\
SDB      &  $ S\times t(b)$ & $b=n^{0.6}$ &  6.810  \\
         &   & $b=n^{0.8}$ &   38.363 \\
Our method & $K\times t(m) + c(K)$ & $K=50$ & 1.031 \\
 &   & $K=100$ & 1.158 \\
&  & $K=150$ & 1.285 \\
\bottomrule
\end{tabular}}
\label{table1}
\end{table}

\section{Simulations}\label{sec3}
In this section, we investigate the finite sample performance of our proposed method. We also compare it with several existing alternatives in the literature. Example \ref{example1} is designed for linear model. Example \ref{example2} is for Logistic regression. Based on the suggestion in \cite{Shi2018}, the numbers of subsets for steps 1 and 2 are 2000 and $10^4$ respectively in  mMSE and mVC. As in \cite{Kleiner2014} and \cite{Sengupta2016}, we set subset size $b=n^{\gamma}$ with $\gamma=0.6$ and $0.8$. The numbers of subsets in BLB and SDB are 20 and 500 respectively. The number of sampled subset is 100 in BLB.   Furthermore, we set the replications of TB to be 100, $K=\{50,100, 150\}$ and $n=10^5$. We  report
empirical sizes and powers for different distributions. Each experiment is repeated 500 times at the nominal level $\alpha=0.05$.

\begin{example}\label{example1}
We consider the linear model: $Y=X^\top\beta + \varepsilon$. Here $\beta$ is a $7\times 1$ vector with all coordinates 0.2 and $X$ comes from the 7-dimensional multivariate normal distribution $N(0, \Sigma)$, where $\Sigma=(\rho_{ij})$ and  $\rho_{ij}=0.2^{|i- j|}$. $\varepsilon$ comes from three distributions:
 \begin{itemize}
 \item[Case 1] The normal distribution, $N(0,1)$.
 \item[Case 2] $t$ distribution, $t$(10).
 \item[Case 3] Mixed normal distribution, $0.5 N(1, 1) + 0.5 N(-1, 1)$.
 \end{itemize}
\end{example}

Table  \ref{table2} shows the empirical sizes when we are testing $H_0: \beta_j=0.2$. Table \ref{table3} summaries the lengths of confidence intervals by different methods.
 We can obtain the following conclusions.
\begin{enumerate}
\item[(1)]   Regardless of distribution of $\varepsilon$, the empirical size of our proposed method outperformes BLB, SDB, and  is slightly better than TB at many cases. Our method is not sensitive to the selection of $K$ since their results are similar.
 \item[(2)]  The empirical sizes of BLB and SDB are close to zero. The possible reason is that the lengths of their confidence intervals are very long, especially when  $\gamma=0.6$.  Compared with BLB and SDB, our method is similar to TB. We also note that in Table \ref{table3}, the lengths in one row are almost the same. This is due to the fact that all $\beta_j$ are set to be equal.
\end{enumerate}

\begin{table}[!h]
\small\renewcommand\arraystretch{0.8}
\centering
\def~{\hphantom{0}}
\caption{Empirical sizes comparison for Example \ref{example1}. }{%
\begin{tabular}{cccc cccc c}
 \\\toprule
Case	&	Method	&	$\beta_{1}$	&	$\beta_{2}$	&	$\beta_{3}$	&	$\beta_{4}$	&	$\beta_{5}$	&	$\beta_{6}$	&	$\beta_7$	\\\midrule
1	&	K=50	&	0.044 	&	0.046 	&	0.046 	&	0.060 	&	0.062 	&	0.056 	&	0.064 	\\
	&	K=100	&	0.060 	&	0.036 	&	0.046 	&	0.054 	&	0.048 	&	0.054 	&	0.074 	\\
	&	K=150	&	0.042 	&	0.034 	&	0.038 	&	0.044 	&	0.044 	&	0.056 	&	0.068 	\\
	&	BLB($n^{0.6}$)	&	0.000 	&	0.000 	&	0.000 	&	0.000 	&	0.000 	&	0.000 	&	0.000 	\\
	&	BLB($n^{0.8}$)	&	0.000 	&	0.000 	&	0.000 	&	0.000 	&	0.000 	&	0.000 	&	0.002 	\\
	&	SDB($n^{0.6}$)	&	0.000 	&	0.000 	&	0.000 	&	0.000 	&	0.000 	&	0.000 	&	0.000 	\\
	&	SDB($n^{0.8}$)	&	0.000 	&	0.000 	&	0.000 	&	0.000 	&	0.000 	&	0.000 	&	0.000 	\\
	&	TB	&	0.068 	&	0.042 	&	0.048 	&	0.074 	&	0.060 	&	0.080 	&	0.076 	\\\midrule
2	&	K=50	&	0.044 	&	0.080 	&	0.038 	&	0.060 	&	0.060 	&	0.060 	&	0.052 	\\
	&	K=100	&	0.032 	&	0.066 	&	0.032 	&	0.054 	&	0.066 	&	0.066 	&	0.046 	\\
	&	K=150	&	0.030 	&	0.054 	&	0.032 	&	0.050 	&	0.064 	&	0.058 	&	0.058 	\\
	&	BLB($n^{0.6}$)	&	0.000 	&	0.000 	&	0.000 	&	0.000 	&	0.000 	&	0.000 	&	0.000 	\\
	&	BLB($n^{0.8}$)	&	0.000 	&	0.002 	&	0.002 	&	0.000 	&	0.000 	&	0.000 	&	0.004 	\\
	&	SDB($n^{0.6}$)	&	0.000 	&	0.000 	&	0.000 	&	0.000 	&	0.000 	&	0.000 	&	0.000 	\\
	&	SDB($n^{0.8}$)	&	0.000 	&	0.000 	&	0.000 	&	0.000 	&	0.000 	&	0.000 	&	0.000 	\\
	&	TB	&	0.044 	&	0.078 	&	0.040 	&	0.064 	&	0.072 	&	0.082 	&	0.080 	\\\midrule
3	&	K=50	&	0.066 	&	0.060 	&	0.042 	&	0.054 	&	0.062 	&	0.052 	&	0.060 	\\
	&	K=100	&	0.054 	&	0.060 	&	0.042 	&	0.062 	&	0.054 	&	0.046 	&	0.046 	\\
	&	K=150	&	0.060 	&	0.056 	&	0.048 	&	0.064 	&	0.054 	&	0.050 	&	0.048 	\\
	&	BLB($n^{0.6}$)	&	0.000 	&	0.000 	&	0.000 	&	0.000 	&	0.000 	&	0.000 	&	0.000 	\\
	&	BLB($n^{0.8}$)	&	0.000 	&	0.002 	&	0.002 	&	0.000 	&	0.002 	&	0.000 	&	0.000 	\\
	&	SDB($n^{0.6}$)	&	0.000 	&	0.000 	&	0.000 	&	0.000 	&	0.000 	&	0.000 	&	0.000 	\\
	&	SDB($n^{0.8}$)	&	0.000 	&	0.000 	&	0.000 	&	0.000 	&	0.000 	&	0.000 	&	0.000 	\\
	&	TB	&	0.070 	&	0.074 	&	0.050 	&	0.082 	&	0.062 	&	0.058 	&	0.066 	\\
\bottomrule
\end{tabular}}
\label{table2}
\end{table}

\begin{table}[!h]
\small\renewcommand\arraystretch{0.8}
\centering
\def~{\hphantom{0}}
\caption{Lengths of confidence interval comparison for Example \ref{example1}. }{%
\begin{tabular}{cccc cccc c}
 \\\toprule
Case	&	Method	&	$\beta_{1}$	&	$\beta_{2}$	&	$\beta_{3}$	&	$\beta_{4}$	&	$\beta_{5}$	&	$\beta_{6}$	&	$\beta_7$	\\\hline
1	&	K=50	&	0.013 	&	0.013 	&	0.013 	&	0.013 	&	0.013 	&	0.013 	&	0.013 	\\
	&	K=100	&	0.013 	&	0.013 	&	0.013 	&	0.013 	&	0.013 	&	0.013 	&	0.013 	\\
	&	K=150	&	0.013 	&	0.013 	&	0.013 	&	0.013 	&	0.013 	&	0.013 	&	0.013 	\\
	&	BLB($n^{0.6}$)	&	0.106 	&	0.107 	&	0.111 	&	0.110 	&	0.109 	&	0.109 	&	0.106 	\\
	&	BLB($n^{0.8}$)	&	0.034 	&	0.035 	&	0.034 	&	0.035 	&	0.035 	&	0.034 	&	0.034 	\\
	&	SDB($n^{0.6}$)	&	0.127 	&	0.129 	&	0.129 	&	0.129 	&	0.129 	&	0.129 	&	0.127 	\\
	&	SDB($n^{0.8}$)	&	0.042 	&	0.042 	&	0.043 	&	0.042 	&	0.042 	&	0.042 	&	0.042 	\\
	&	TB	&	0.012 	&	0.012 	&	0.012 	&	0.012 	&	0.012 	&	0.012 	&	0.012 	\\\midrule
2	&	K=50	&	0.014 	&	0.014 	&	0.014 	&	0.014 	&	0.014 	&	0.014 	&	0.014 	\\
	&	K=100	&	0.014 	&	0.015 	&	0.015 	&	0.015 	&	0.015 	&	0.014 	&	0.014 	\\
	&	K=150	&	0.014 	&	0.015 	&	0.014 	&	0.015 	&	0.014 	&	0.015 	&	0.014 	\\
	&	BLB($n^{0.6}$)	&	0.120 	&	0.122 	&	0.122 	&	0.121 	&	0.122 	&	0.123 	&	0.118 	\\
	&	BLB($n^{0.8}$)	&	0.037 	&	0.039 	&	0.038 	&	0.038 	&	0.038 	&	0.038 	&	0.037 	\\
	&	SDB($n^{0.6}$)	&	0.142 	&	0.144 	&	0.144 	&	0.144 	&	0.144 	&	0.145 	&	0.141 	\\
	&	SDB($n^{0.8}$)	&	0.046 	&	0.047 	&	0.047 	&	0.047 	&	0.048 	&	0.048 	&	0.046 	\\
	&	TB	&	0.014 	&	0.014 	&	0.014 	&	0.014 	&	0.014 	&	0.014 	&	0.014 	\\\midrule
3	&	K=50	&	0.018 	&	0.018 	&	0.018 	&	0.018 	&	0.018 	&	0.018 	&	0.018 	\\
	&	K=100	&	0.018 	&	0.018 	&	0.018 	&	0.018 	&	0.018 	&	0.018 	&	0.018 	\\
	&	K=150	&	0.018 	&	0.018 	&	0.018 	&	0.018 	&	0.018 	&	0.018 	&	0.018 	\\
	&	BLB($n^{0.6}$)	&	0.150 	&	0.155 	&	0.155 	&	0.154 	&	0.155 	&	0.153 	&	0.150 	\\
	&	BLB($n^{0.8}$)	&	0.047 	&	0.048 	&	0.049 	&	0.049 	&	0.049 	&	0.048 	&	0.047 	\\
	&	SDB($n^{0.6}$)	&	0.179 	&	0.182 	&	0.183 	&	0.182 	&	0.182 	&	0.182 	&	0.178 	\\
	&	SDB($n^{0.8}$)	&	0.059 	&	0.060 	&	0.060 	&	0.060 	&	0.060 	&	0.060 	&	0.059 	\\
	&	TB	&	0.017 	&	0.018 	&	0.017 	&	0.018 	&	0.017 	&	0.017 	&	0.017 	\\
\bottomrule
\end{tabular}}
\label{table3}
\end{table}

\clearpage
\begin{example}\label{example2}
In this example,
 we consider a $p$-dimensional  multiple Logistic regression model. Given covariates $Z_{i}\in \R^{p}$,
$$
\P(Y_{i}=1|Z_{i})=\frac{\exp(Z_{i}^\top \beta)}{1+ \exp(Z_{i}^\top \beta)}, \quad i=1,\dots, n,
$$
where $Y_{i}\in\{0,1\}$ is the response and $\beta$ is a $p$-dimensional unknown parameter.
The interesting problem is to test the hypothesis:  $\beta_{j}=\beta_{j0}$ for some $1\leq j\leq p$, or $\beta=\beta_{0}$.

 We let $\beta$ be a $7\times 1$ vector with all coordinates equal to 0.2. $Z_i$ comes from seven distributions which were used in \cite{Shi2018}.
\begin{itemize}
  \item[Case 1] $N(0, \Sigma)$, $\Sigma=(\rho_{ij})$  with $\rho_{ij}=0.5^{I(i\neq j)}$, where $I(\cdot)$ is the indicator function.
  \item[Case 2] $N(1.5, \Sigma)$.
  \item[Case 3] $0.5 N(1, \Sigma) + 0.5 N(-1, \Sigma)$.
  \item[Case 4] The multivariate $t$ distribution $t_{3}(0, \Sigma)/10$, with degrees of freedom 3.
  \item[Case 5] The multivariate exponential distribution whose components are independent and each has
an exponential distribution with a rate parameter of 2.
  \item[Case 6] $0.5 N(-2.14, \Sigma) + 0.5 N(-2.9, \Sigma)$.
\end{itemize}
Here, Cases 2 and 5 produce imbalanced data. Case 6 produces rare events data.
\end{example}

Tables  \ref{table4}-\ref{table7} show the empirical sizes and powers. When we consider powers of test, the null hypothesis is that the parameter $\beta_j$ is zero.  Tables \ref{table8} and \ref{table9} summarize the lengths of confidence intervals.  We draw the following conclusions.
\begin{enumerate}
\item[(1)]   Regardless of  imbalanced data or the rare events data, the empirical sizes of our proposed  method are close to the nominal level, which implies our method  performs well. Moreover, the empirical power is very close to 1. The differences among three values of K is not significantly.
 \item[(2)] As  $\gamma$ increases, the performance of BLB and SDB becomes better. However, they are worse than our method. TB slightly inflated rejection probabilities under the null hypothesis. From Tables \ref{table8} and \ref{table9}, the length of confidence intervals  decreases as  $\gamma$  increases. When $\gamma=0.6$, it is ten times as long as TB, which results  in the lower empirical size.
 \item[(3)] In terms of  empirical powers, mVC and mMSE  outperform  BLB and SDB. Compared with  mVC and mMSE, our method is better, especially in the case imbalanced data and rare events data in terms of empirical sizes and powers.
\end{enumerate}

\begin{table}[!h]
\small\renewcommand\arraystretch{0.6}
\centering
\def~{\hphantom{0}}
\caption{Empirical sizes comparison for Cases 1-3 in Example \ref{example2}. }{%
\begin{tabular}{cccc cccc c}
 \\\toprule
Case	&	Method	&	$\beta_{1}$	&	$\beta_{2}$	&	$\beta_{3}$	&	$\beta_{4}$	&	$\beta_{5}$	&	$\beta_{6}$	&	$\beta_7$	\\\midrule
1 	&	K=50	&	0.054 	&	0.064 	&	0.052 	&	0.052 	&	0.038 	&	0.040 	&	0.048 	\\
	&	K=100	&	0.056 	&	0.062 	&	0.044 	&	0.050 	&	0.050 	&	0.038 	&	0.056 	\\
	&	K=150	&	0.070 	&	0.060 	&	0.062 	&	0.050 	&	0.052 	&	0.050 	&	0.048 	\\
	&	mVC	&	0.056 	&	0.076 	&	0.058 	&	0.044 	&	0.038 	&	0.086 	&	0.072 	\\
	&	mMSE	&	0.068 	&	0.046 	&	0.068 	&	0.062 	&	0.074 	&	0.078 	&	0.044 	\\
	&	BLB($n^{0.6}$)	&	0.000 	&	0.000 	&	0.000 	&	0.000 	&	0.000 	&	0.002 	&	0.000 	\\
	&	BLB($n^{0.8}$)	&	0.004 	&	0.002 	&	0.000 	&	0.004 	&	0.000 	&	0.000 	&	0.000 	\\
	&	SDB($n^{0.6}$)	&	0.000 	&	0.000 	&	0.000 	&	0.000 	&	0.000 	&	0.000 	&	0.000 	\\
	&	SDB($n^{0.8}$)	&	0.000 	&	0.000 	&	0.000 	&	0.000 	&	0.000 	&	0.000 	&	0.000 	\\
	&	TB	&	0.066 	&	0.068 	&	0.058 	&	0.056 	&	0.042 	&	0.042 	&	0.072 	\\\midrule
2 	&	K=50	&	0.044 	&	0.050 	&	0.054 	&	0.058 	&	0.066 	&	0.052 	&	0.060 	\\
	&	K=100	&	0.050 	&	0.046 	&	0.050 	&	0.060 	&	0.054 	&	0.032 	&	0.052 	\\
	&	K=150	&	0.054 	&	0.052 	&	0.060 	&	0.070 	&	0.054 	&	0.040 	&	0.052 	\\
	&	mVC	&	0.076 	&	0.098 	&	0.092 	&	0.086 	&	0.094 	&	0.078 	&	0.070 	\\
	&	mMSE	&	0.076 	&	0.086 	&	0.086 	&	0.074 	&	0.082 	&	0.098 	&	0.060 	\\
	&	BLB($n^{0.6}$)	&	0.000 	&	0.000 	&	0.000 	&	0.000 	&	0.000 	&	0.000 	&	0.000 	\\
	&	BLB($n^{0.8}$)	&	0.002 	&	0.000 	&	0.000 	&	0.004 	&	0.002 	&	0.000 	&	0.000 	\\
	&	SDB($n^{0.6}$)	&	0.000 	&	0.000 	&	0.000 	&	0.000 	&	0.000 	&	0.000 	&	0.000 	\\
	&	SDB($n^{0.8}$)	&	0.000 	&	0.000 	&	0.000 	&	0.000 	&	0.000 	&	0.000 	&	0.000 	\\
	&	TB	&	0.062 	&	0.054 	&	0.064 	&	0.076 	&	0.074 	&	0.050 	&	0.070 	\\\midrule
3 	&	K=50	&	0.060 	&	0.054 	&	0.058 	&	0.040 	&	0.040 	&	0.040 	&	0.058 	\\
	&	K=100	&	0.060 	&	0.048 	&	0.048 	&	0.036 	&	0.032 	&	0.046 	&	0.058 	\\
	&	K=150	&	0.074 	&	0.044 	&	0.050 	&	0.054 	&	0.040 	&	0.046 	&	0.056 	\\
	&	mVC	&	0.054 	&	0.064 	&	0.084 	&	0.070 	&	0.058 	&	0.066 	&	0.058 	\\
	&	mMSE	&	0.066 	&	0.078 	&	0.048 	&	0.066 	&	0.084 	&	0.072 	&	0.084 	\\
	&	BLB($n^{0.6}$)	&	0.000 	&	0.000 	&	0.000 	&	0.000 	&	0.000 	&	0.000 	&	0.000 	\\
	&	BLB($n^{0.8}$)	&	0.000 	&	0.000 	&	0.000 	&	0.002 	&	0.000 	&	0.000 	&	0.002 	\\
	&	SDB($n^{0.6}$)	&	0.000 	&	0.000 	&	0.000 	&	0.000 	&	0.000 	&	0.000 	&	0.000 	\\
	&	SDB($n^{0.8}$)	&	0.000 	&	0.000 	&	0.000 	&	0.000 	&	0.000 	&	0.000 	&	0.000 	\\
	&	TB	&	0.080 	&	0.054 	&	0.076 	&	0.060 	&	0.054 	&	0.050 	&	0.066 	\\
\bottomrule
\end{tabular}}
\label{table4}
\end{table}

\begin{table}[!h]
\small\renewcommand\arraystretch{0.6}
\centering
\def~{\hphantom{0}}
\caption{Empirical sizes comparison for Cases 4-6 in Example \ref{example2}. }{%
\begin{tabular}{cccc cccc c}
 \\\toprule
Case	&	Method	&	$\beta_{1}$	&	$\beta_{2}$	&	$\beta_{3}$	&	$\beta_{4}$	&	$\beta_{5}$	&	$\beta_{6}$	&	$\beta_7$	\\\hline
4 	&	K=50	&	0.058 	&	0.058 	&	0.058 	&	0.048 	&	0.062 	&	0.068 	&	0.062 	\\
	&	K=100	&	0.056 	&	0.062 	&	0.058 	&	0.054 	&	0.052 	&	0.054 	&	0.044 	\\
	&	K=150	&	0.046 	&	0.062 	&	0.058 	&	0.054 	&	0.058 	&	0.054 	&	0.060 	\\
	&	mVC	&	0.078 	&	0.072 	&	0.070 	&	0.080 	&	0.054 	&	0.070 	&	0.066 	\\
	&	mMSE	&	0.070 	&	0.080 	&	0.058 	&	0.068 	&	0.060 	&	0.066 	&	0.068 	\\
	&	BLB($n^{0.6}$)	&	0.000 	&	0.000 	&	0.000 	&	0.000 	&	0.000 	&	0.000 	&	0.000 	\\
	&	BLB($n^{0.8}$)	&	0.000 	&	0.000 	&	0.000 	&	0.000 	&	0.002 	&	0.002 	&	0.000 	\\
	&	SDB($n^{0.6}$)	&	0.000 	&	0.000 	&	0.000 	&	0.000 	&	0.000 	&	0.000 	&	0.000 	\\
	&	SDB($n^{0.8}$)	&	0.000 	&	0.000 	&	0.000 	&	0.000 	&	0.000 	&	0.000 	&	0.000 	\\
	&	TB	&	0.070 	&	0.082 	&	0.062 	&	0.074 	&	0.064 	&	0.060 	&	0.084 	\\\midrule
5 	&	K=50	&	0.066 	&	0.062 	&	0.046 	&	0.074 	&	0.046 	&	0.058 	&	0.048 	\\
	&	K=100	&	0.056 	&	0.058 	&	0.054 	&	0.070 	&	0.048 	&	0.066 	&	0.050 	\\
	&	K=150	&	0.060 	&	0.060 	&	0.066 	&	0.084 	&	0.044 	&	0.068 	&	0.052 	\\
	&	mVC	&	0.060 	&	0.082 	&	0.066 	&	0.090 	&	0.070 	&	0.068 	&	0.066 	\\
	&	mMSE	&	0.074 	&	0.074 	&	0.048 	&	0.068 	&	0.052 	&	0.070 	&	0.064 	\\
	&	BLB($n^{0.6}$)	&	0.000 	&	0.000 	&	0.000 	&	0.000 	&	0.000 	&	0.000 	&	0.000 	\\
	&	BLB($n^{0.8}$)	&	0.000 	&	0.000 	&	0.000 	&	0.002 	&	0.000 	&	0.000 	&	0.002 	\\
	&	SDB($n^{0.6}$)	&	0.000 	&	0.000 	&	0.000 	&	0.000 	&	0.000 	&	0.000 	&	0.000 	\\
	&	SDB($n^{0.8}$)	&	0.000 	&	0.000 	&	0.000 	&	0.000 	&	0.000 	&	0.000 	&	0.000 	\\
	&	TB	&	0.062 	&	0.070 	&	0.074 	&	0.070 	&	0.060 	&	0.074 	&	0.064 	\\\midrule
6 	&	K=50	&	0.064 	&	0.070 	&	0.048 	&	0.050 	&	0.070 	&	0.050 	&	0.066 	\\
	&	K=100	&	0.060 	&	0.078 	&	0.048 	&	0.068 	&	0.058 	&	0.042 	&	0.052 	\\
	&	K=150	&	0.062 	&	0.074 	&	0.056 	&	0.062 	&	0.056 	&	0.044 	&	0.060 	\\
	&	mVC	&	0.126 	&	0.160 	&	0.132 	&	0.154 	&	0.134 	&	0.152 	&	0.124 	\\
	&	mMSE	&	0.140 	&	0.152 	&	0.146 	&	0.146 	&	0.154 	&	0.142 	&	0.152 	\\
	&	BLB($n^{0.6}$)	&	0.000 	&	0.000 	&	0.000 	&	0.000 	&	0.000 	&	0.000 	&	0.000 	\\
	&	BLB($n^{0.8}$)	&	0.000 	&	0.002 	&	0.000 	&	0.000 	&	0.000 	&	0.000 	&	0.000 	\\
	&	SDB($n^{0.6}$)	&	0.000 	&	0.000 	&	0.000 	&	0.000 	&	0.000 	&	0.000 	&	0.000 	\\
	&	SDB($n^{0.8}$)	&	0.000 	&	0.000 	&	0.000 	&	0.000 	&	0.000 	&	0.000 	&	0.000 	\\
	&	TB	&	0.078 	&	0.082 	&	0.068 	&	0.076 	&	0.076 	&	0.054 	&	0.070 	\\
\bottomrule
\end{tabular}}
\label{table5}
\end{table}

\begin{table}[!h]
\small\renewcommand\arraystretch{0.6}
\centering
\def~{\hphantom{0}}
\caption{Empirical powers comparison for Cases 1-3 in Example \ref{example2}. }{%
\begin{tabular}{cccc cccc c}
 \\\toprule
Case	&	Method	&	$\beta_{1}$	&	$\beta_{2}$	&	$\beta_{3}$	&	$\beta_{4}$	&	$\beta_{5}$	&	$\beta_{6}$	&	$\beta_7$	\\\hline
1 	&	K=50	&	1.000 	&	1.000 	&	1.000 	&	1.000 	&	1.000 	&	1.000 	&	1.000 	\\
	&	K=100	&	1.000 	&	1.000 	&	1.000 	&	1.000 	&	1.000 	&	1.000 	&	1.000 	\\
	&	K=150	&	1.000 	&	1.000 	&	1.000 	&	1.000 	&	1.000 	&	1.000 	&	1.000 	\\
	&	mVC	&	1.000 	&	1.000 	&	1.000 	&	1.000 	&	1.000 	&	1.000 	&	1.000 	\\
	&	mMSE	&	1.000 	&	1.000 	&	1.000 	&	1.000 	&	1.000 	&	1.000 	&	1.000 	\\
	&	BLB($n^{0.6}$)	&	0.884 	&	0.894 	&	0.848 	&	0.896 	&	0.864 	&	0.872 	&	0.880 	\\
	&	BLB($n^{0.8}$)	&	1.000 	&	1.000 	&	1.000 	&	1.000 	&	1.000 	&	1.000 	&	1.000 	\\
	&	SDB($n^{0.6}$)	&	0.910 	&	0.900 	&	0.910 	&	0.908 	&	0.876 	&	0.900 	&	0.884 	\\
	&	SDB($n^{0.8}$)	&	1.000 	&	1.000 	&	1.000 	&	1.000 	&	1.000 	&	1.000 	&	1.000 	\\
	&	TB	&	1.000 	&	1.000 	&	1.000 	&	1.000 	&	1.000 	&	1.000 	&	1.000 	\\\midrule
2 	&	K=50	&	1.000 	&	1.000 	&	1.000 	&	1.000 	&	1.000 	&	1.000 	&	1.000 	\\
	&	K=100	&	1.000 	&	1.000 	&	1.000 	&	1.000 	&	1.000 	&	1.000 	&	1.000 	\\
	&	K=150	&	1.000 	&	1.000 	&	1.000 	&	1.000 	&	1.000 	&	1.000 	&	1.000 	\\
	&	mVC	&	1.000 	&	1.000 	&	1.000 	&	1.000 	&	1.000 	&	1.000 	&	1.000 	\\
	&	mMSE	&	1.000 	&	1.000 	&	1.000 	&	1.000 	&	1.000 	&	1.000 	&	1.000 	\\
	&	BLB($n^{0.6}$)	&	0.464 	&	0.472 	&	0.428 	&	0.458 	&	0.480 	&	0.478 	&	0.488 	\\
	&	BLB($n^{0.8}$)	&	1.000 	&	1.000 	&	1.000 	&	1.000 	&	1.000 	&	1.000 	&	1.000 	\\
	&	SDB($n^{0.6}$)	&	0.006 	&	0.010 	&	0.006 	&	0.008 	&	0.010 	&	0.004 	&	0.008 	\\
	&	SDB($n^{0.8}$)	&	1.000 	&	1.000 	&	1.000 	&	1.000 	&	1.000 	&	1.000 	&	1.000 	\\
	&	TB	&	1.000 	&	1.000 	&	1.000 	&	1.000 	&	1.000 	&	1.000 	&	1.000 	\\\midrule
3 	&	K=50	&	1.000 	&	1.000 	&	1.000 	&	1.000 	&	1.000 	&	1.000 	&	1.000 	\\
	&	K=100	&	1.000 	&	1.000 	&	1.000 	&	1.000 	&	1.000 	&	1.000 	&	1.000 	\\
	&	K=150	&	1.000 	&	1.000 	&	1.000 	&	1.000 	&	1.000 	&	1.000 	&	1.000 	\\
	&	mVC	&	0.944 	&	0.960 	&	0.944 	&	0.966 	&	0.950 	&	0.960 	&	0.968 	\\
	&	mMSE	&	0.976 	&	0.976 	&	0.978 	&	0.976 	&	0.952 	&	0.986 	&	0.962 	\\
	&	BLB($n^{0.6}$)	&	0.030 	&	0.070 	&	0.042 	&	0.066 	&	0.042 	&	0.044 	&	0.032 	\\
	&	BLB($n^{0.8}$)	&	0.994 	&	0.988 	&	0.998 	&	0.998 	&	0.994 	&	0.996 	&	0.994 	\\
	&	SDB($n^{0.6}$)	&	0.000 	&	0.000 	&	0.000 	&	0.000 	&	0.000 	&	0.000 	&	0.000 	\\
	&	SDB($n^{0.8}$)	&	0.998 	&	1.000 	&	1.000 	&	1.000 	&	0.998 	&	1.000 	&	0.998 	\\
	&	TB	&	1.000 	&	1.000 	&	1.000 	&	1.000 	&	1.000 	&	1.000 	&	1.000 	\\
\bottomrule
\end{tabular}}
\label{table6}
\end{table}

\begin{table}[!h]
\small\renewcommand\arraystretch{0.6}
\centering
\def~{\hphantom{0}}
\caption{Empirical powers comparison for Cases 4-6 in Example \ref{example2}. }{%
\begin{tabular}{cccc cccc c}
 \\\toprule
Case	&	Method	&	$\beta_{1}$	&	$\beta_{2}$	&	$\beta_{3}$	&	$\beta_{4}$	&	$\beta_{5}$	&	$\beta_{6}$	&	$\beta_7$	\\\hline
4	&	K=50	&	0.976 	&	0.982 	&	0.972 	&	0.962 	&	0.972 	&	0.962 	&	0.964 	\\
	&	K=100	&	0.972 	&	0.970 	&	0.964 	&	0.956 	&	0.978 	&	0.960 	&	0.950 	\\
	&	K=150	&	0.974 	&	0.976 	&	0.956 	&	0.956 	&	0.962 	&	0.956 	&	0.966 	\\
	&	mVC	&	0.354 	&	0.388 	&	0.340 	&	0.364 	&	0.366 	&	0.356 	&	0.372 	\\
	&	mMSE	&	0.400 	&	0.352 	&	0.384 	&	0.416 	&	0.368 	&	0.356 	&	0.376 	\\
	&	BLB($n^{0.6}$)	&	0.000 	&	0.000 	&	0.000 	&	0.000 	&	0.002 	&	0.000 	&	0.002 	\\
	&	BLB($n^{0.8}$)	&	0.206 	&	0.230 	&	0.224 	&	0.230 	&	0.210 	&	0.226 	&	0.230 	\\
	&	SDB($n^{0.6}$)	&	0.000 	&	0.000 	&	0.000 	&	0.000 	&	0.000 	&	0.000 	&	0.000 	\\
	&	SDB($n^{0.8}$)	&	0.014 	&	0.018 	&	0.006 	&	0.016 	&	0.010 	&	0.012 	&	0.006 	\\
	&	TB	&	0.982 	&	0.980 	&	0.984 	&	0.968 	&	0.986 	&	0.972 	&	0.970 	\\\midrule
5 	&	K=50	&	1.000 	&	1.000 	&	1.000 	&	1.000 	&	1.000 	&	1.000 	&	1.000 	\\
	&	K=100	&	1.000 	&	1.000 	&	1.000 	&	1.000 	&	1.000 	&	1.000 	&	1.000 	\\
	&	K=150	&	1.000 	&	1.000 	&	1.000 	&	1.000 	&	1.000 	&	1.000 	&	1.000 	\\
	&	mVC	&	1.000 	&	0.998 	&	1.000 	&	1.000 	&	0.998 	&	1.000 	&	1.000 	\\
	&	mMSE	&	1.000 	&	1.000 	&	1.000 	&	1.000 	&	1.000 	&	1.000 	&	1.000 	\\
	&	BLB($n^{0.6}$)	&	0.394 	&	0.418 	&	0.406 	&	0.442 	&	0.424 	&	0.466 	&	0.394 	\\
	&	BLB($n^{0.8}$)	&	1.000 	&	1.000 	&	1.000 	&	1.000 	&	1.000 	&	1.000 	&	1.000 	\\
	&	SDB($n^{0.6}$)	&	0.002 	&	0.000 	&	0.012 	&	0.000 	&	0.010 	&	0.004 	&	0.004 	\\
	&	SDB($n^{0.8}$)	&	1.000 	&	1.000 	&	1.000 	&	1.000 	&	1.000 	&	1.000 	&	1.000 	\\
	&	TB	&	1.000 	&	1.000 	&	1.000 	&	1.000 	&	1.000 	&	1.000 	&	1.000 	\\\midrule
6 	&	K=50	&	0.996 	&	0.996 	&	0.998 	&	0.984 	&	0.994 	&	0.996 	&	0.994 	\\
	&	K=100	&	0.994 	&	0.998 	&	0.998 	&	0.990 	&	0.988 	&	0.994 	&	0.996 	\\
	&	K=150	&	0.994 	&	1.000 	&	0.998 	&	0.986 	&	0.992 	&	0.998 	&	0.990 	\\
	&	mVC	&	0.918 	&	0.922 	&	0.944 	&	0.930 	&	0.958 	&	0.922 	&	0.958 	\\
	&	mMSE	&	0.942 	&	0.962 	&	0.954 	&	0.944 	&	0.950 	&	0.954 	&	0.958 	\\
	&	BLB($n^{0.6}$)	&	0.000 	&	0.000 	&	0.006 	&	0.000 	&	0.002 	&	0.000 	&	0.000 	\\
	&	BLB($n^{0.8}$)	&	0.312 	&	0.310 	&	0.358 	&	0.370 	&	0.346 	&	0.324 	&	0.354 	\\
	&	SDB($n^{0.6}$)	&	0.000 	&	0.000 	&	0.000 	&	0.000 	&	0.000 	&	0.000 	&	0.000 	\\
	&	SDB($n^{0.8}$)	&	0.020 	&	0.030 	&	0.040 	&	0.022 	&	0.042 	&	0.040 	&	0.030 	\\
	&	TB	&	0.994 	&	0.998 	&	0.996 	&	0.992 	&	0.994 	&	0.996 	&	0.992 	\\
\bottomrule
\end{tabular}}
\label{table7}
\end{table}

\begin{table}[!h]
\small\renewcommand\arraystretch{0.6}
\centering
\def~{\hphantom{0}}
\caption{ Lengths of confidence interval for Cases 1-3 in Example \ref{example2} }{%
\begin{tabular}{cccc cccc c}
 \\\toprule
Case	&	Method	&	$\beta_{1}$	&	$\beta_{2}$	&	$\beta_{3}$	&	$\beta_{4}$	&	$\beta_{5}$	&	$\beta_{6}$	&	$\beta_7$	\\\hline
1 	&	K=50	&	0.037 	&	0.037 	&	0.037 	&	0.037 	&	0.037 	&	0.037 	&	0.037 	\\
	&	K=100	&	0.037 	&	0.037 	&	0.037 	&	0.037 	&	0.037 	&	0.037 	&	0.037 	\\
	&	K=150	&	0.038 	&	0.038 	&	0.037 	&	0.037 	&	0.037 	&	0.037 	&	0.037 	\\
	&	mVC	&	0.048 	&	0.048 	&	0.048 	&	0.048 	&	0.048 	&	0.048 	&	0.048 	\\
	&	mMSE	&	0.047 	&	0.047 	&	0.047 	&	0.047 	&	0.047 	&	0.047 	&	0.047 	\\
	&	BLB($n^{0.6}$)	&	0.313 	&	0.313 	&	0.319 	&	0.311 	&	0.311 	&	0.316 	&	0.313 	\\
	&	BLB($n^{0.8}$)	&	0.097 	&	0.096 	&	0.098 	&	0.096 	&	0.097 	&	0.097 	&	0.098 	\\
	&	SDB($n^{0.6}$)	&	0.370 	&	0.370 	&	0.370 	&	0.370 	&	0.369 	&	0.370 	&	0.369 	\\
	&	SDB($n^{0.8}$)	&	0.120 	&	0.120 	&	0.120 	&	0.121 	&	0.121 	&	0.121 	&	0.120 	\\
	&	TB	&	0.035 	&	0.035 	&	0.035 	&	0.035 	&	0.035 	&	0.035 	&	0.035 	\\ \midrule
2 	&	K=50	&	0.050 	&	0.049 	&	0.049 	&	0.049 	&	0.049 	&	0.050 	&	0.049 	\\
	&	K=100	&	0.050 	&	0.050 	&	0.050 	&	0.050 	&	0.050 	&	0.050 	&	0.050 	\\
	&	K=150	&	0.051 	&	0.051 	&	0.051 	&	0.051 	&	0.051 	&	0.051 	&	0.051 	\\
	&	mVC	&	0.050 	&	0.050 	&	0.050 	&	0.050 	&	0.050 	&	0.050 	&	0.050 	\\
	&	mMSE	&	0.047 	&	0.047 	&	0.047 	&	0.047 	&	0.047 	&	0.047 	&	0.047 	\\
	&	BLB($n^{0.6}$)	&	0.416 	&	0.424 	&	0.423 	&	0.422 	&	0.419 	&	0.424 	&	0.423 	\\
	&	BLB($n^{0.8}$)	&	0.131 	&	0.133 	&	0.130 	&	0.132 	&	0.132 	&	0.133 	&	0.132 	\\
	&	SDB($n^{0.6}$)	&	0.503 	&	0.502 	&	0.500 	&	0.500 	&	0.499 	&	0.500 	&	0.501 	\\
	&	SDB($n^{0.8}$)	&	0.162 	&	0.162 	&	0.162 	&	0.162 	&	0.162 	&	0.162 	&	0.162 	\\
	&	TB	&	0.047 	&	0.047 	&	0.047 	&	0.047 	&	0.047 	&	0.047 	&	0.047 	\\\midrule
3 	&	K=50	&	0.082 	&	0.082 	&	0.082 	&	0.082 	&	0.082 	&	0.082 	&	0.082 	\\
	&	K=100	&	0.083 	&	0.083 	&	0.083 	&	0.083 	&	0.083 	&	0.083 	&	0.083 	\\
	&	K=150	&	0.084 	&	0.084 	&	0.084 	&	0.083 	&	0.083 	&	0.084 	&	0.083 	\\
	&	mVC	&	0.103 	&	0.103 	&	0.103 	&	0.103 	&	0.103 	&	0.103 	&	0.103 	\\
	&	mMSE	&	0.096 	&	0.095 	&	0.096 	&	0.096 	&	0.096 	&	0.096 	&	0.095 	\\
	&	BLB($n^{0.6}$)	&	0.707 	&	0.702 	&	0.697 	&	0.696 	&	0.686 	&	0.701 	&	0.690 	\\
	&	BLB($n^{0.8}$)	&	0.218 	&	0.219 	&	0.217 	&	0.217 	&	0.219 	&	0.214 	&	0.216 	\\
	&	SDB($n^{0.6}$)	&	0.827 	&	0.826 	&	0.827 	&	0.824 	&	0.825 	&	0.824 	&	0.825 	\\
	&	SDB($n^{0.8}$)	&	0.270 	&	0.268 	&	0.268 	&	0.269 	&	0.268 	&	0.268 	&	0.269 	\\
	&	TB	&	0.078 	&	0.078 	&	0.078 	&	0.078 	&	0.078 	&	0.078 	&	0.078 	\\
\bottomrule
\end{tabular}}
\label{table8}
\end{table}

\begin{table}[!h]
\small\renewcommand\arraystretch{0.6}
\centering
\def~{\hphantom{0}}
\caption{ Lengths of confidence interval for Cases 4-6 in Example \ref{example2} }{%
\begin{tabular}{cccc cccc c}
 \\\toprule
Case	&	Method	&	$\beta_{1}$	&	$\beta_{2}$	&	$\beta_{3}$	&	$\beta_{4}$	&	$\beta_{5}$	&	$\beta_{6}$	&	$\beta_7$	\\\hline
4 	&	K=50	&	0.207 	&	0.207 	&	0.208 	&	0.207 	&	0.206 	&	0.206 	&	0.206 	\\
	&	K=100	&	0.213 	&	0.213 	&	0.213 	&	0.214 	&	0.213 	&	0.213 	&	0.212 	\\
	&	K=150	&	0.219 	&	0.218 	&	0.218 	&	0.218 	&	0.218 	&	0.218 	&	0.218 	\\
	&	mVC	&	0.250 	&	0.249 	&	0.250 	&	0.250 	&	0.250 	&	0.250 	&	0.250 	\\
	&	mMSE	&	0.240 	&	0.240 	&	0.241 	&	0.240 	&	0.241 	&	0.241 	&	0.240 	\\
	&	BLB($n^{0.6}$)	&	1.793 	&	1.790 	&	1.810 	&	1.785 	&	1.800 	&	1.772 	&	1.792 	\\
	&	BLB($n^{0.8}$)	&	0.531 	&	0.541 	&	0.539 	&	0.534 	&	0.531 	&	0.540 	&	0.538 	\\
	&	SDB($n^{0.6}$)	&	2.129 	&	2.118 	&	2.119 	&	2.128 	&	2.128 	&	2.128 	&	2.131 	\\
	&	SDB($n^{0.8}$)	&	0.662 	&	0.663 	&	0.661 	&	0.662 	&	0.663 	&	0.663 	&	0.661 	\\
	&	TB	&	0.191 	&	0.191 	&	0.192 	&	0.190 	&	0.192 	&	0.190 	&	0.190 	\\\midrule
5 	&	K=50	&	0.052 	&	0.052 	&	0.052 	&	0.052 	&	0.052 	&	0.052 	&	0.052 	\\
	&	K=100	&	0.052 	&	0.052 	&	0.052 	&	0.052 	&	0.052 	&	0.052 	&	0.052 	\\
	&	K=150	&	0.053 	&	0.053 	&	0.053 	&	0.053 	&	0.053 	&	0.053 	&	0.053 	\\
	&	mVC	&	0.065 	&	0.065 	&	0.065 	&	0.065 	&	0.065 	&	0.065 	&	0.065 	\\
	&	mMSE	&	0.064 	&	0.064 	&	0.064 	&	0.064 	&	0.064 	&	0.064 	&	0.064 	\\
	&	BLB($n^{0.6}$)	&	0.441 	&	0.435 	&	0.441 	&	0.440 	&	0.442 	&	0.439 	&	0.430 	\\
	&	BLB($n^{0.8}$)	&	0.138 	&	0.136 	&	0.138 	&	0.139 	&	0.138 	&	0.136 	&	0.138 	\\
	&	SDB($n^{0.6}$)	&	0.523 	&	0.523 	&	0.522 	&	0.521 	&	0.523 	&	0.522 	&	0.523 	\\
	&	SDB($n^{0.8}$)	&	0.170 	&	0.169 	&	0.169 	&	0.170 	&	0.169 	&	0.169 	&	0.169 	\\
	&	TB	&	0.049 	&	0.049 	&	0.049 	&	0.049 	&	0.049 	&	0.049 	&	0.049 	\\\midrule
6 	&	K=50	&	0.178 	&	0.178 	&	0.178 	&	0.178 	&	0.178 	&	0.178 	&	0.179 	\\
	&	K=100	&	0.181 	&	0.181 	&	0.182 	&	0.182 	&	0.182 	&	0.182 	&	0.182 	\\
	&	K=150	&	0.185 	&	0.186 	&	0.185 	&	0.185 	&	0.186 	&	0.185 	&	0.185 	\\
	&	mVC	&	0.098 	&	0.098 	&	0.098 	&	0.098 	&	0.098 	&	0.098 	&	0.098 	\\
	&	mMSE	&	0.093 	&	0.093 	&	0.093 	&	0.093 	&	0.093 	&	0.093 	&	0.093 	\\
	&	BLB($n^{0.6}$)	&	1.521 	&	1.512 	&	1.538 	&	1.516 	&	1.522 	&	1.530 	&	1.526 	\\
	&	BLB($n^{0.8}$)	&	0.466 	&	0.466 	&	0.472 	&	0.459 	&	0.463 	&	0.467 	&	0.468 	\\
	&	SDB($n^{0.6}$)	&	1.806 	&	1.815 	&	1.803 	&	1.813 	&	1.816 	&	1.806 	&	1.811 	\\
	&	SDB($n^{0.8}$)	&	0.579 	&	0.578 	&	0.576 	&	0.578 	&	0.579 	&	0.577 	&	0.579 	\\
	&	TB	&	0.168 	&	0.167 	&	0.168 	&	0.169 	&	0.167 	&	0.168 	&	0.167 	\\
\bottomrule
\end{tabular}}
\label{table9}
\end{table}

\clearpage
\section{A real data} \label{sec4}
In this section, we apply the proposed method to a census income data set, which aims to determine whether a person makes \$50K or more a year. The  data can be obtained from  \url{https://archive.ics.uci.edu/ml/datasets/census+income}, with 48,842 observations in total. As in \cite{Wang2018},  the response
variable is whether a person's income exceeds \$50K a year. The  explanatory variables are as follows:
\begin{itemize}
\setlength{\itemsep}{0pt}
\setlength{\parsep}{0pt}
\setlength{\parskip}{0pt}
  \item[]$X_{1}$: age
  \item[]$X_{2}$: final weight (Fnlwgt)
  \item[]$X_{3}$: highest level of education in numerical form (Education-num)
  \item[]$X_{4}$: capital loss (Capital-loss);
  \item[]$X_{5}$: hours worked per week (Hours-per-week).
\end{itemize}
There are 11,687 individuals
(23.929\%) in the data whose income exceeds \$50K a year.  In order to eliminate the effect of scale, we have scaled and centered each explanatory variable so that they have mean 0 and variance 1.
To evaluate the performance of the above methods, we replicate each method 500 times since these methods split  sample  randomly. We report the average  estimate and the average  proportion of rejecting the null hypothesis that the regression coefficient is zero by all methods.

Table \ref{table10} shows the result.   The traditional Logistic regression (TLR) indicates that all coefficients are significant, not equal to 0  under the nominal level 5\%. Our method is consistent to  the traditional Logistic regression. Compared with $K=150$ and $K=50$, $K=100$ is better since each block sample contains enough data points.  For $\beta_{3}$,  the average  proportion of rejecting the null hypothesis by mVC, mMSE, BLB and SDB are much lower than 1 while ours are 1. It implies that our proposed method works in cases where others don't work.

\begin{table}[!h]
\small\renewcommand\arraystretch{0.9}
\centering
\def~{\hphantom{0}}
\caption{ The results of a real data. }{%
\begin{tabular}{cccc ccc}
 \\\toprule
Method &	$\beta_{1}$	&	$\beta_{2}$	&	$\beta_{3}$	&	$\beta_{4}$	&	$\beta_{5}$	&	$\beta_{6}$	\\\midrule
\multicolumn{7}{c}{ Estimate } \\ \midrule
TLR	&	-1.514 	&	0.630 	&	0.063 	&	0.877 	&	0.226 	&	0.521 	\\\midrule
\multicolumn{7}{c}{ Average Estimate } \\\midrule
K=50	&	-1.525 	&	0.637 	&	0.063 	&	0.885 	&	0.229 	&	0.529 	\\
K=100	&	-1.537 	&	0.644 	&	0.063 	&	0.896 	&	0.231 	&	0.538 	\\
K=150	&	-1.549 	&	0.651 	&	0.062 	&	0.905 	&	0.234 	&	0.547 	\\
mVC	&	-1.510 	&	0.627 	&	0.066 	&	0.876 	&	0.225 	&	0.527 	\\
mMSE	&	-1.514 	&	0.634 	&	0.059 	&	0.876 	&	0.229 	&	0.518 	\\\midrule

\multicolumn{7}{c}{$p$-value } \\ \hline
TLR		&	0.000 	&	0.000 	&	0.000 	&	0.000 	&	0.000 	&	0.000 	\\\hline
\multicolumn{7}{c}{ Average  Proportion } \\ \hline
K=50	&	1.000 	&	1.000 	&	1.000 	&	1.000 	&	1.000 	&	1.000 	\\
K=100	&	1.000 	&	1.000 	&	1.000 	&	1.000 	&	1.000 	&	1.000 	\\
K=150	&	1.000 	&	1.000 	&	1.000 	&	1.000 	&	0.990 	&	1.000 	\\
mVC	&	1.000 	&	1.000 	&	0.860 	&	1.000 	&	1.000 	&	1.000 	\\
mMSE	&	1.000 	&	1.000 	&	0.770 	&	1.000 	&	1.000 	&	1.000 	\\
BLB($n^{0.6}$)	&	1.000 	&	1.000 	&	0.000 	&	1.000 	&	0.910 	&	1.000 	\\
BLB($n^{0.8}$)	&	1.000 	&	1.000 	&	0.750 	&	1.000 	&	1.000 	&	1.000 	\\
SDB($n^{0.6}$)	&	1.000 	&	1.000 	&	0.000 	&	1.000 	&	1.000 	&	1.000 	\\
SDB($n^{0.8}$)	&	1.000 	&	1.000 	&	0.010 	&	1.000 	&	1.000 	&	1.000 	\\
\bottomrule
\end{tabular}}
\label{table10}
\end{table}

\clearpage

\appendix
\renewcommand{\theequation}{A.\arabic{equation}}
\setcounter{equation}{0}
\section*{Appendix}

{\bf Proof of Theorem \ref{theorem1}}.

From Assumption \ref{assumption1}, we can get
$$
\sqrt{m} (\hat{\theta}_{km}-\theta )=\frac{1}{\sqrt{m}}\sum_{i=1}^{m}\eta_{ki}+R_{km}.
$$
Hence,
\begin{align}\label{eqA1}
 \sqrt{n}(\widetilde{\theta} - \theta )  & = \sqrt{n} \Big{(}\frac{1}{K}\sum_{k=1}^{K} \hat{\theta}_{km} -\theta \Big{)} \notag \\
   &=  \frac{\sqrt{n}}{K}\sum_{k=1}^{K} (\hat{\theta}_{km} -\theta ) \notag \\
   &=  \frac{1}{\sqrt{K}}\sum_{k=1}^{K} \sqrt{m} (\hat{\theta}_{km} -\theta ) \notag \\
   &=\frac{1}{\sqrt{K}}\sum_{k=1}^{K}  W_{km}+ \frac{1}{\sqrt{K}}\sum_{k=1}^{K} R_{km},
\end{align}
where $W_{km}=\frac{1}{\sqrt{m}}\sum_{i=1}^{m}\eta_{k,i}$. 
From Assumption \ref{assumption2}, we get the last term in (\ref{eqA1}) is $o_{p}(1)$.

 Now, we prove that $\frac{1}{\sqrt{K}}\sum_{k=1}^{K} W_{km} $ has the asymptotic normality distribution. Let $V_{km}=c^\top W_{km}$, then  $\E(V_{km})=0, Var(V_{km})=c^\top \Sigma c=\sigma^2$. By the Cram\'{e}r-Wold theorem, we only need to prove
$$
  \frac{1}{\sqrt{K}}\sum_{k=1}^{K} V_{km} \stackrel{d}{\longrightarrow}N(0,\sigma^2)
$$
for each fixed $c\in\mathbb{R}^p\setminus\{0\}$.

Since $V_{km}$ is a normalized sum of $K$  independent and identically distributed random variables, it follows from Linderberg's CLT that
$$
 \E e^{\imath tV_{mk}(u)}=e^{-t^2\sigma^2/2}+o(t^2),
$$
for any real $t\in \R$. Here $\imath=\sqrt{-1}$.

Hence,
\begin{align*}
 &\E\exp\left\{ \imath t\frac{1}{\sqrt{K}}\sum_{k=1}^K V_{km}\right\}\\
  &=\left(\E e^{\imath tK^{-1/2} V_{km}}\right)^K\\
&=\Big(e^{-(tK^{-1/2})^2\sigma^2/2}+o(tK^{-1/2})^2\Big)^K\to e^{-t^2\sigma^2/2},
\end{align*}
as $K\to\infty$.  The proof of Theorem \ref{theorem1} is completed.

For proving Theorem \ref{theorem2}, we need the following two lemmas.
\begin{lemma}\label{lem-1}
Let $Z_{K}=\max_{1\leq k\leq K} \|Y_{km}- \mu \|$. Under the conditions of Theorem \ref{theorem2}, we have
$$
Z_{K}=o_p(K^{1/2})
$$
as $K, m\to \infty$.
\end{lemma}
\proof Note that
$$
Y_{km} - \mu =\frac{1}{\sqrt{m}}\sum_{i=1}^m\eta_{ki}+R_{km}.
$$
Since $\eta_{ki}$'s are independent and identically distributed random vectors with mean zero and finite fourth moment,
\begin{align*}
\P\Big(\max_{1\leq k\leq K} \Big\|\frac{1}{\sqrt{m}}\sum_{i=1}^m\eta_{ki}\Big\|>\epsilon \sqrt{K}\Big)
&\leq \sum_{k=1}^K \P\Big(  \Big\|\frac{1}{\sqrt{m}}\sum_{i=1}^m\eta_{ki}\Big\|>\epsilon \sqrt{K} \Big) \\
&\leq K ( \epsilon \sqrt{K})^{-4}  \E\Big\|\frac{1}{\sqrt{m}}\sum_{i=1}^m\eta_{ki}\Big\|^4\\
&=O(K^{-1})
\end{align*}
as $K, m\to \infty$. On the other hand,  from the Assumption \ref{assumption2},  we get
$$
 \P\Big(\max_{1\leq i\leq K} \|R_{km}\|>\epsilon K^{1/2}\Big)\to 0
$$
as $K, m\to \infty$. So we can complete the proof.

\begin{lemma}\label{lem-2}
Let
$$
  S_K =\frac{1}{K}\sum_{k=1}^{K}(Y_{km}-\mu)(Y_{km}-\mu)^\top.
$$
 Under the conditions of Theorem \ref{theorem2}, we have $S_K \stackrel{p}{\longrightarrow}  \Sigma $ as $K,m\to \infty$.
\end{lemma}
\proof Note that
\begin{align*}
   & (Y_{km}-\mu)(Y_{km}-\mu)^\top \\
   &=\Big(\frac{1}{\sqrt{m}}\sum_{i=1}^m\eta_{ki}+R_{km}\Big)\Big(\frac{1}{\sqrt{m}}\sum_{i=1}^m\eta_{ki}+R_{km}\Big)^\top\\
   &=\frac{1}{m}\Big(\sum_{i=1}^m\eta_{ki}\Big)\Big(\sum_{i=1}^m\eta_{ki}\Big)^\top
+2\frac{1}{\sqrt{m}}\Big(\sum_{i=1}^m\eta_{ki}\Big)R_{km}^\top + R_{km}R_{km}^\top.
\end{align*}
Now we consider the convergence of the $(j,l)$ element of $S_K$ for $1\leq j,l\leq p$.
For any $\epsilon>0$,
\begin{align*}
 & \P\Big(K^{-1}\Big|\frac{1}{\sqrt{m}}\sum_{k=1}^K (\sum_{i=1}^m\eta_{kij})R_{kml}\Big| > \epsilon\Big)\\
   &\leq  \P\Big(\max_{1\leq k\leq K}\Big| \Big(\sum_{i=1}^m\eta_{kij}\Big)R_{kml}\Big| > \sqrt{m}\epsilon\Big)  \\
   &\leq   \sum_{k=1}^K \P\Big(| \sum_{i=1}^m\eta_{kij}| >C m^{1/2+\alpha} \Big) +  \P\Big(\max_{1\leq k\leq K}|R_{kml}| >C^{-1}m^{-\alpha} \epsilon \Big),
\end{align*}
where $C$ is a constant which will go to infinity finally. Since $\eta_{kij}$, $k,i=1,2,\cdots$ are  independent and identically distributed random variables with mean zero and finite fourth moment,
$$ \P\Big(| \sum_{i=1}^m\eta_{kij}| > Cm^{1/2+\alpha}\Big)\leq C^{-4}m^{-2-4\alpha} \E| \sum_{i=1}^m\eta_{kij}|^4 =C^{-4}O(m^{-4\alpha}).$$
It follows from Assumption \ref{assumption2} that
$$\P\Big(K^{-1}|\sum_{k=1}^K (\sum_{i=1}^m\eta_{kij})R_{kml}| > \epsilon\Big)  \to 0,$$
if we let $K,m\to \infty$ as a first step, then let $C\to \infty$ as a second step. Similarly,
$$\P\Big(K^{-1}|\sum_{k=1}^K R_{kmj}R_{kml}|>\epsilon\Big)\to 0$$
as $K,m\to \infty$. It remains to consider
\begin{equation} \label{eqA2}
(Km)^{-1} \sum_{k=1}^K\Big(\sum_{i=1}^m \eta_{kij}\Big)\Big(\sum_{i=1}^m \eta_{kil}\Big)=(Km)^{-1} \sum_{k=1}^K\sum_{i=1}^m \eta_{kij}\eta_{kil}+(Km)^{-1} \sum_{k=1}^K\sum_{1\leq i_1\not=i_2\leq m}\eta_{ki_1j}\eta_{ki_2l}.
\end{equation}
The second sum on the right hand side of equality in \eqref{eqA2} converges to zero in probability by Markov's inequality as $K,m\to \infty$. The first sum  on the right hand side of equality in \eqref{eqA2} converges to the $(j,l)$ element of $\Sigma$ in probability by law of large numbers. Combining all above completes the proof.

\noindent{\bf Proof Theorem \ref{theorem2}}.
 (\ref{eq24}) can be re-expression as
\begin{equation}\label{eqA3}
f(\lambda)=\frac{1}{K}\sum_{k=1}^{K} \frac{Y_{km}- \mu}{1 + \lambda^\top(Y_{km}- \mu)}=0.
\end{equation}
 Let $\lambda=\| \lambda\|\theta$, where $\theta\in \Theta$ is a unit vector, and $\Theta$ denotes the set of unit vector in $\mathbb{R}^p$. In the following, we show
 $$
\|\lambda\|= O_{p}(K^{-1/2}).
$$

Let
 $$
 U_{km}= \lambda^\top(Y_{km}- \mu).
 $$
Using the representation $1/(1+ U_{km})=1 - U_{m,k}/ (1+U_{km})$, and  $\theta^\top f(\lambda)=0$, we have
\begin{equation}\label{eqA4}
  \theta^\top (\bar{Y}_{Km}- \mu) =\| \lambda \| \theta^\top \tilde{S} \theta,
\end{equation}
where
$$
\tilde{S}=\frac{1}{K}\sum_{k=1}^{K}\frac{(Y_{km}-\mu)(Y_{km}-\mu)^\top}{1+ U_{km}}
$$
and
$$
\bar{Y}_{Km} = \frac{1}{K} \sum_{k=1}^{K}Y_{km}.
$$
Since $0<\omega_{k}<1$, we have $1+U_{m, k} >0$, hence
\begin{align*}
  \|\lambda\|\theta^\top S_K \theta  & \leq \|\lambda\|\theta^\top \tilde{S} \theta (1+ \max_{1\leq k\leq K} U_{km})\\
  &\leq \|\lambda\|\theta^\top \tilde{S} \theta (1+ \|\lambda\| Z_{K})\\
  &=  \theta^\top (\bar{Y}_{Km}- \mu)(1+ \|\lambda\| Z_{K}).
\end{align*}
The last equality follows by (\ref{eqA4}). Hence,
$$
 \|\lambda\|[ \theta^\top S_K\theta -  \theta^\top(\bar{Y}_{Km}-\mu) Z_{K}] \leq   \theta^\top (\bar{Y}_{Km}-\mu).
$$
By the central limit theorem, $\bar{Y}_{Km}-\mu=O_{p}(K^{-1/2})$.  Lemma \ref{lem-1} shows $Z_{K}=o_{p}(K^{1/2})$. By Lemma \ref{lem-2}, the smallest eigenvalue of $S$ always has a positive lower bound in probability. Combing these three facts, it gives
$$
 \|\lambda\|[ \theta^\top S_K\theta + O_{p}(K^{-1/2}) o_{p}(K^{1/2}) ] = O_{p}(K^{-1/2}).
$$
So, we have
$$
\|\lambda\|= O_{p}(K^{-1/2}).
$$
Furthermore,
\begin{equation}\label{eqA5}
  \max_{1\leq k\leq K}|U_{km}|= O_{p}(K^{-1/2})o_{p}(K^{-1/2})=o_{p}(1).
\end{equation}
Expanding (\ref{eqA3}) gives
\begin{align}\label{eqA6}
   0&=\frac{1}{K}\sum_{k=1}^{n} (Y_{km}-  \mu) \Big{\{} 1 - U_{km} + \frac{U_{km}^{2}}{1 + U_{km}}   \Big{\}} \notag \\
    &=(\bar{Y}_{km}-  \mu) - S_K \lambda + \frac{1}{K}\sum_{k=1}^{K} \frac{(Y_{km}-  \mu)U_{km}^{2}}{1 + U_{km}}.
\end{align}
The final term in (\ref{eqA6}) above has a norm bounded by
$$
 \frac{1}{K}\sum_{k=1}^{K}\|Y_{km}- \mu \| ^{3} \|\lambda\| ^2 |1+Y_{km}|^{-1} = o_{p}(K^{1/2})O_{p}(K^{-1}) O_{p}(1)=o_{p}(K^{-1/2}).
$$
So,
$$
\lambda = S_K^{-1}(\bar{Y}_{km}-\mu) + \beta,
$$
with $\beta =o_{p}(K^{-1/2})$. By (\ref{eqA5}), we may expand
$$
\log\Big{(}  1+ U_{m,k} \Big{)}= U_{m, k} - \frac{1}{2}U_{m,k}^{2} + \eta_{k}
$$
where for some finite $B >0, 1\leq k\leq K$,
$$
\P(| \eta_{k}|\leq B|U_{km}|^3 )\to 1
$$
as $K\to \infty$ and $m\to \infty$.

We can verify the follow the identities after some algebra
\begin{align*}
  -2 \log \mathcal{R}(\mu) &=2\sum_{k=1}^{K}\log \Big{(}  1+ U_{km} \Big{)} \\
  &= 2\sum_{k=1}^{K} \Big{(}  U_{km} - \frac{1}{2}U_{km}^{2} + \eta_{k} \Big{)} \\
   &=2K\lambda^\top(\bar{Y}_{Km} -\mu) -K\lambda^\top S_K\lambda + 2 \sum_{k=1}^{K} \eta_{i}\\
   &= K(\bar{Y}_{Km} -\mu)^\top S_K^{-1} (\bar{Y}_{Km} -\mu) -K\beta^\top S_K^{-1}\beta + 2 \sum_{k=1}^{K} \eta_{k}.
\end{align*}
By Theorem \ref{theorem1}  and Lemma \ref{lem-2}
$$
K(\bar{Y}_{km}-\mu)^\top S_K^{-1} (\bar{Y}_{km}-\mu) \stackrel{d}{\longrightarrow} \chi^{2}_{p}.
$$
The second and third terms are $o_{p}(1)$ since
$$
K\beta^\top S_K^{-1}\beta=Ko_{p}(K^{-1/2})O_{p}(1)o_{p}(K^{-1/2})=o_{p}(1),
$$
$$
\Big{|}   \sum_{k=1}^{K} \eta_{k}  \Big{|}\leq B\|\lambda\|^3 \sum_{k=1}^{K} \| Y_{km}-\mu\|^{3} =O_{p}(K^{-3/2})o_{p}(K^{3/2})=o_{p}(1).
$$
Combing above, we can finish the proof.

\end{document}